# Surface plasmon polaritons in the ultraviolet region

E. D. Chubchev,[1] I. A. Nechepurenko,[1] A. V. Dorofeenko,[1,2] A. P. Vinogradov,[1,2] A. A. Lisyansky,[3,4]

[1]*Dukhov Research Institute of Automatics, 22 Suschevskaya, Moscow 120755, Russia*
[2]*Institute for Theoretical and Applied Electromagnetics of Russian Academy of Sciences, 13 Izhorskaya, Moscow 125412, Russia*
[3]*Department of Physics, Queens College of the City University of New York, Queens, NY 11367, USA*
[4]*The Graduate Center of the City University of New York, New York, NY 10016, USA*

We study a surface plasmon polariton mode that is strongly confined in the transverse direction and propagates along a periodically nanostructured metal-dielectric interface. We show that the wavelength of this mode is determined by the period of the structure, and may therefore, be orders of magnitude smaller than the wavelength of a plasmon-polariton propagating along a flat surface. This plasmon polariton exists in the frequency region in which the sum of the real parts of the permittivities of the metal and dielectric is positive, a frequency region in which surface plasmon polaritons do not exist on a flat surface. The propagation length of the new mode can reach a several dozen wavelengths. This mode can be observed in materials that are uncommon in plasmonics, such as aluminum or sodium.

## 1. INTRODUCTION

Our appetite for higher-speed devices inevitably leads to the transition from electronic or optoelectronic to all-optical devices. At the same time, the necessity for higher clock frequencies for information processing requires greater integration of photonic devices and scaling them down to nanometers. In optics, dielectric fiber replaces the coaxial and strip transmission lines. However, the characteristic transverse size of a fiber line is orders of magnitude larger than the characteristic size of components of semiconductor integrated circuitry. Moreover, the radius of curvature for optical line bending reaches hundreds of microns. The resulting large total size of the device hinders the realization of a high clock frequency which is limited by the signal propagation time within the device. A possible solution for this problem is a transition from photons to surface plasmon-polaritons (SPPs).

The SPP is an electromagnetic wave propagating along the interface between a metal and dielectric. The wavelength of an SPP, $\lambda_{SPP} = 2\pi / \operatorname{Re} k_{SPP}$, is smaller than that of the electromagnetic wave in the dielectric with permittivity $\varepsilon_d$, $\lambda_0 / \sqrt{\varepsilon_d}$. The SPP is therefore confined to the metal surface with the transverse (perpendicular to the surface) confinement length $\delta = \lambda_0 \lambda_{SPP} \left( 2\pi (\lambda_0^2 - \varepsilon_d \lambda_{SPP}^2)^{1/2} \right)^{-1}$. In the absence of losses, the SPP propagation length, $l_{pr} = 1/\left( 2 \operatorname{Im} k_{SPP} \right)$, would be infinite and both $\lambda_{SPP}$ and $\delta$ would tend to zero as $\varepsilon_m(\omega)$ tends to $-\varepsilon_d$. For $\lambda_{SPP} \ll \lambda_0$, the confinement length is $\delta \sim \lambda_{SPP}/2\pi \ll \lambda_0$. This enables the miniaturization of optical devices and the transition from electronics to on-chip plasmonics technology possible [1-6].

The main obstacle to the use of SPPs in applications is ohmic loss in the metal. This substantially decreases the SPP propagation length, $l_{pr}$ and also raises the minimum SPP wavelength, $\lambda_{SPP}^{cutoff}$. SPPs with $\lambda_{SPP}$ smaller than $\lambda_{SPP}^{cutoff}$ cannot exist [1]. This weakens the transverse confinement of the SPP. The values of $l_{pr}$, $\lambda_{SPP}^{cutoff}$, and $\delta$ are strongly determined by the geometrical configuration.

In the simplest geometrical configuration of an SPP propagating along the flat interface between half-spaces filled by the metal and dielectric, the wavenumber of the SPP is given by

$$k_{SPP} = k_0\sqrt{\varepsilon_m \varepsilon_d/(\varepsilon_m + \varepsilon_d)}, \qquad (1)$$

where $k_0$ is the wavenumber of the electromagnetic wave in free space [1]. At low frequencies ($|\varepsilon_m(\omega)| \gg 1$), the SPP wavelength is of the order of the wavelength in the dielectric ($\lambda_{SPP} \sim \lambda_0/\sqrt{\varepsilon_d}$) and the transverse confinement length is much greater than $\lambda_{SPP}$. In lossy media, according to Eq. (1), the $\lambda_{SPP}$ decreases as the frequency increases and reaches the minimum non-zero value, $\lambda_{SPP}^{cutoff}$, (see Fig. 1a, curve 1). Thus that there is a minimum transverse confinement length $\delta_{min}$. Cutoff frequencies for silver-vacuum and gold-vacuum interfaces are $\lambda_{SPP}^{cutoff} \approx 0.78\lambda_0$ and $\lambda_{SPP}^{cutoff} \approx 0.91\lambda_0$, respectively. Minimal transverse confinement lengths of SPPs in silver and gold are about $0.2\lambda_0$ and $0.36\lambda_0$, respectively. Unfortunately, at the maximal confinement ($\delta = \delta_{min}$), an SPP cannot propagate at all since $l_{pr}^{cutoff} < \lambda_{SPP}^{cutoff}$.

The SPP propagation length can be increased by shifting into the low-frequency part of the spectrum. The decrease in the frequency corresponds to the movement along curve 1 in Fig. 1a starting from the point $A_1$. Note that both $\lambda_{SPP}$ and $l_{pr}$ increase with decreasing frequency, but $l_{pr}$ increases more rapidly. Therefore, a decrease in the frequency corresponds to a movement to the right along the horizontal axis in Fig. 1b. As curve 1 in Fig. 1b shows, the transverse confinement length increases with a decrease in frequency (this curve starts at the point $A_1$ which corresponds to the plasmon dispersion curve crossing the light cone).

The propagation length may also be increased, even to hundreds of wavelengths, by using thin metal films [7]. The corresponding dispersion curve for long-range SPPs is determined by the equation

$$\frac{\varepsilon_m}{\varepsilon_d} = i\frac{\sqrt{\varepsilon_m k_0^2 - k_{SPP}^2}}{\sqrt{\varepsilon_d k_0^2 - k_{SPP}^2}} \tan\left(\sqrt{\varepsilon_m k_0^2 - k_{SPP}^2}\,\frac{h}{2}\right), \qquad (2)$$

where $h$ is the thickness of the metal film. According to Eq. (2), the dispersion curve of the long-range SPP is close to the light cone (see curve 2 in Fig. 1a), so its wavelength $\lambda_{SPP}$ is about $\lambda_0$, while the confinement length tends towards infinity [7-9]. In other words, the SPP tends to a plane wave, and the plasmonic thin films turn into an analog of a single-wire transmittance line (see curve 2 in Figs. 1a, b). At the point $A_2$, curve 2 ends crossing the line cone. Such a transmittance line has no advantages compared to a common optical dielectric waveguide.

Using chains of metal nanoparticles allows one to decrease both the wavelength of an SPP and its transverse confinement length [10-16]. If the dipole moment of a nanoparticle is parallel to the axis of the chain, then in the quasistatic and tight binding approximation, the dispersion law of such a chain is

$$k_{SPP} = \frac{1}{a}\arccos\left[\frac{a^3}{4\alpha(\omega)}\right], \qquad (3)$$

where $\alpha(\omega) = [\varepsilon_m(\omega) - \varepsilon_d]/[\varepsilon_m(\omega) + 2\varepsilon_d]r^3$ is the polarizability of a metal nanosphere of radius $r$ [17]. The shortest wavelength is of the order of the distance between nanoparticles, $a$. Since the SPP frequency is close to the plasmon resonance of the nanoparticle, in the quasistatic approach, the electric field is mainly concentrated inside the nanoparticles [18]. Therefore, losses in such chains are large, and the propagation length of an SPP is only a few $\lambda_{SPP}$ (in Figs. 1a, b, curve 3 originates and ends at the points $A_3$ and $B_3$ that are the points at which the dispersion curve crosses the light cone). Even though the SPP confinement length in a chain may be small (of the order of $a$), the propagation length is negligibly short.

Thus, in the visible region, even the best plasmonic materials, such as gold and silver, are not suitable for applications that require $\delta \ll \lambda_0$ and $l_{pr}$ of the order of a dozen SPP wavelengths. Crossing from optical to the infrared region can significantly improve the characteristics of the system. This has been observed in chains of strongly elongated plasmonic particles [19] (see curve 3 in Fig. 1b). For example, in a chain of spheroids with the period of 20 nm and the ration of semiaxes of 0.15, the propagation length is about $700\lambda_{SPP}$.

The search for new plasmonic materials has been growing sharply in recent years [20-23]. In particular, the transition to the ultraviolet part of the spectrum brings into consideration new materials, such as aluminum, sodium, and rubidium [22,24]. It allows for significant enhancements of fluorescence and of the rates of photochemical reactions [24-27]. In Sections II–IV, using aluminum as example, we consider propagation of strongly localized SPPs. The effects under consideration cannot be observed in silver and gold. It would also be highly desirable to obtain topologically more complicated interfaces that combine advantages of the aforementioned systems without their shortcomings.

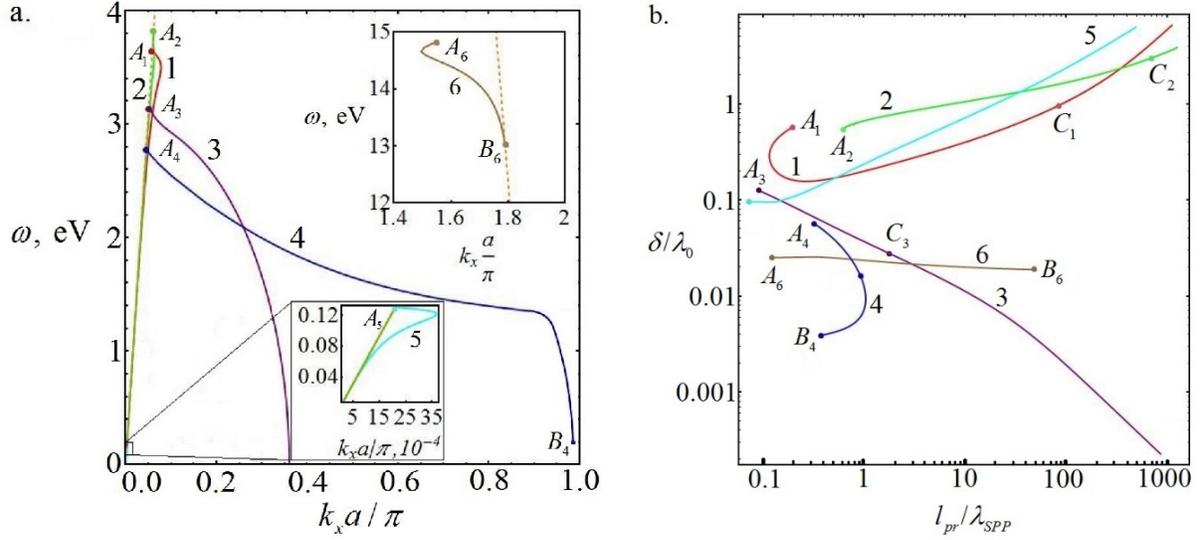

Fig. 1. The dispersion curves (a) and the parametric dependence of the transverse confinement length $\delta(\omega)$ on the propagation length $l_{pr}(\omega)$ (b) for different topologies that support SPPs. The red line 1 corresponds to an SPP propagating on a flat surface of silver half-space, the green line 2 corresponds to a long-range SPP propagating along a silver 30-nm-thin film. In this systems, the unlimited growth of $\delta$ is due to the transition to the IR region. The purple 3, blue 4, and brown 6 lines show propagation lengths of SPPs along the chain of silver nanospheres, the nanostructured silver-vacuum interface, and nanostructured aluminum-vacuum interface, respectively. Cyan line 5 corresponds to the spoof. Curves 3 and 5 have been calculated by using formulas from Refs. [19] and [28], respectively. Curves 1-5 have been calculated for values of metal permittivity taken from Ref. [29], while curve 6 is calculated using the data from Ref. [30]. In Fig. 1b, points $C_i$ correspond to the boundary of the visible region assumed to be 780 nm; segments $A_iB_i$ correspond to the visible region. For numerical calculations in this figure and in the manuscript, we assume that the dielectric is vacuum with $\varepsilon_d = 1$.

In this paper, we demonstrate that a periodically nanostructured metal-dielectric interface (an array of metal nanoparticles deposited on the metal surface) supports an SPP mode with a short transverse confinement length that is comparable to the period of the structure (see curves 4 and 5 in Figs. 1a, b). In contrast to spoof SPPs that exist on structured surfaces in the frequency range in which $|\varepsilon_m(\omega)| \gg \varepsilon_d$ [28,31-33] (curve 5 in Fig. 1), the eigenfrequency of this mode is in the ultraviolet part of the spectrum in which $-\varepsilon_d < \operatorname{Re}\varepsilon_m(\omega) < 0$. This is far from the usual plasmonic resonances ensuring a longer propagation length. In addition, the wavelength of the SPP is approximately the same as the period of the surface structure. This differs from a spoof SPP which wavelength belongs to the first Brillouin zone near the light cone. The SPP that we consider has a subwavelength confinement thanks to its small wavelength. This SPP mode cannot be observed in traditional plasmonic materials due to high losses caused by interband transitions in this part of the spectrum. In aluminum, the required frequencies are in the ultraviolet region, in which losses are relatively small because interband transitions are in the visible region. Therefore, the propagation length of the SPP can be as large as hundreds of nanometers on a periodically nanostructured aluminum-dielectric interface.

## 2. NANOSTRUCTURED SURFACE

In the absence of losses, on a rough metal surface, an additional SPP may arise [34]. The solution corresponding to this SPP has been obtained with the assumption that for a smooth surface, near the frequency for which $\varepsilon_m(\omega) = -\varepsilon_d$, the group velocity of the SPP is zero. This provides a resonance interaction of the field with all harmonics of the roughness. As a result, the frequency curve splits and the second SPP mode with large $k$ arises. However, in a lossy system, on a smooth surface, an SPP with sufficiently large wavenumbers does not exist. In addition, the assumption of zero group velocity, which is necessary for the second branch of the SPP, is not realistic. Therefore, it is not clear whether in a lossy system an additional SPP may arise.

To model periodically nanostructured metal-dielectric interface, we consider a system shown in Fig. 2. The interface between the metal and dielectric is modulated by the function $s(x) = h\cos(2\pi x/a)$. The upper half-space is a dielectric with the permittivity $\varepsilon_d$ and the lower half-space is a metal with the permittivity $\varepsilon_m$. We look for a frequency-domain solution as a TM-polarized wave propagating along the x-axis. Since the roughness is periodic, we seek for the solution in the form $H_y = F(x,z)e^{ik_x x}$, where $H_y$ is the magnetic field strength of the SPP, $F(x,z)$ is a periodic in the x-direction function with the period $a$, and $k_x = k_x' + ik_x''$ is the propagation constant.

To solve Maxwell's equations, we make the coordinate transformation [35,36] which makes the interface flat, but coefficients in the equations become periodic with respect to $x$. This method is applicable for a large amplitude of roughness when the Rayleigh hypothesis is invalid [37,38]. Within the framework of this method, the tangential components of the electric and magnetic fields are represented as a series of Bloch harmonics. This allows one to reduce Maxwell's equations to the system of linear differential equations with constant coefficients and to obtain expressions for the fields in both media analytically. By using the Maxwell boundary conditions, one can obtain the propagation constant $k_x(\omega)$ of the eigensolution as a function of frequency.

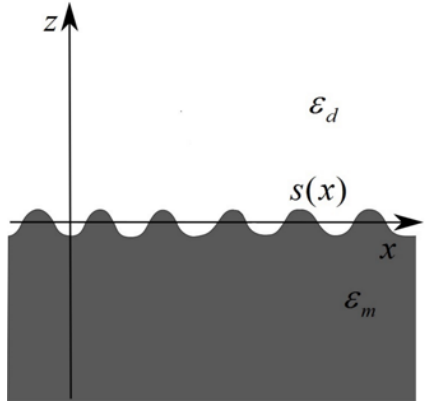

Fig. 2. The schematics of the system studied.

## 3. ANALYSIS OF THE DISPERSION CURVES FOR THE NANOSTRUCTURED SURFACE

For an SPP, a periodically nanostructured surface is a photonic crystal. For typical plasmonic materials such as silver and gold, the SPP curve in the second Brillouin zone cannot be observed due to high losses in the UV frequency region in which these losses are due to the interband transitions. In the UV part of the spectrum of aluminum, losses are small because interband transitions are in the visible part of the spectrum. Therefore, in aluminum, this SPP dispersion branch can be observed.

First, we consider a lossless vacuum-aluminum system. The dispersion functions, $k_x(\omega)$, of SPPs on the nanostructured surface for various amplitudes of the interface profile are shown in Fig. 3. The dispersion curve $k_x(\omega)$ of the SPP propagating on the flat surface ($h = 0$) is described by the well-known formula [1]: $k_x = k_0 \left[ \varepsilon_m \varepsilon_v / (\varepsilon_m + \varepsilon_v) \right]^{1/2}$ (shown by the blue line in Fig. 3). For small $h$, similar to photonic crystals, due to the Bragg reflection of plasmon-polaritons between neighboring surface inhomogeneities, the band gap at $k_x a / \pi = 1$ opens up (the rad line). The width of this band gap increases with an increase of $h$ [39]. At a certain value of $h$, in the second Brillouin zone, a part of the dispersion curve moves to the frequency region, in which $\operatorname{Re}\varepsilon_m(\omega) > -\varepsilon_d$. In this region, SPPs on a flat metal-vacuum interface do not exist because $k_x$ becomes pure imaginary. One might expect that the "plasmon" band gap should arise in this region. However, there is a pass band because, in a periodic system, the energy can be transferred by evanescent fields [11].

In a lossless system, the dispersion curves for interfaces with technologically achievable amplitudes of the surface perturbation ($h = 5$ nm in our calculations) are shown in Fig. 4a. In this case, the band gap arises for the wavelengths 97 nm – 207 nm. In the first Brillouin zone, the dispersion curve of the SPP becomes non-monotonic, and a point in which the group velocity is zero arises. In the second Brillouin zone, the dispersion curve moves completely to the frequency region defined by the inequality $\operatorname{Re}\varepsilon_m(\omega) > -\varepsilon_d$. In this zone, the waves are backward.

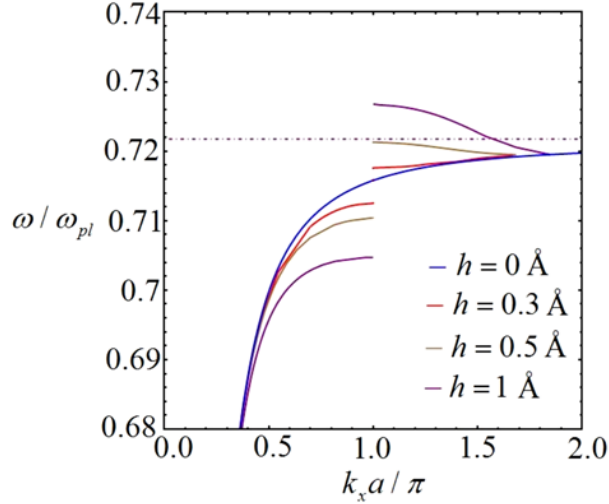

Fig. 3. The dispersion curves of the SPP on the nanostructured surface for various amplitudes $h$. The period of the modulation is 10 nm. The dielectric permittivity of metal assumed to be equal to the permittivity of aluminum taken from Ref. [30]. The dash-dotted purple line corresponds to $\omega/\omega_{pl} = 0.73$ at which $\mathrm{Re}\varepsilon_m = -\varepsilon_d$.

Losses change the SPP dispersion curves significantly (see curve 4 and 5 in Figs. 1 a, b and 4b). One can see that in the first Brillouin zone, at the point in which in the absence of loss, the group velocity is zero, the SPP curve splits into two branches (curves 4 and 4'). Both of these branches are in the visible region. One of the branches (curve 4) has a negative slope corresponding to the backward wave. This wave exists far from the light cone where it may be strongly confined. Because the respective wavenumber is of the order of $\pi/2a$ (see Fig. 3b). However, computer simulations show that on an aluminum surface, the propagation length of the SPP associated with this curve is small. It is not more than the SPP wavelength. Even for silver, which in visible does not have interband transitions, the propagation length does not exceed a wavelength of the SPP. The dependence of the confinement length of this mode in silver is shown by curve 4 in Fig. 1b.

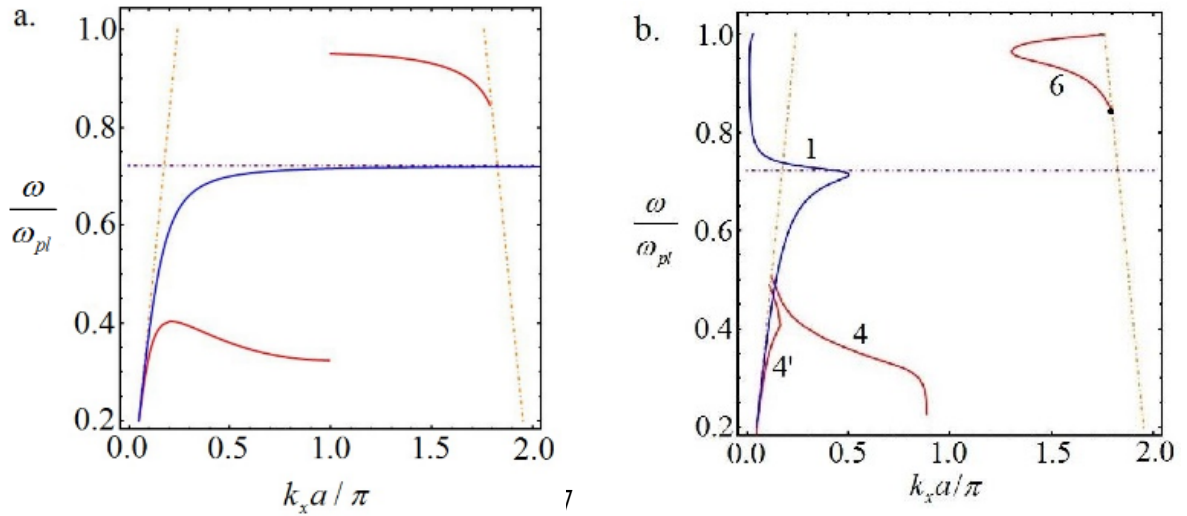

Fig. 4. The dispersion curves of the SPP on an aluminum surface calculated without (*a*) and with (*b*) taking loss into account. The dispersion curve of the SPP on flat and nanostructured surfaces are shown by blue and red lines, respectively. The structure parameters are $h = 5$ nm and $a = 10$ nm. Orange lines show boundaries of the light cone, the dash-dotted purple line corresponds to $\omega/\omega_{pl} = 0.73$. $\mathrm{Re}\,\varepsilon_m = -\varepsilon_d$ for the frequency $\omega = 0.73\omega_{pl}$. In this figure, as well as in Fig. 5, the line numbering is the same as in Fig. 1.

Losses cause significant distortion of the dispersion curve in the second Brillouin zone. For typical plasmonic materials such as silver and gold, the SPP curve in the second Brillouin zone cannot be observed due to high losses in the UV frequency region in which these losses are due to the interband transitions. In aluminum, interband transitions are in the visible part of the spectrum, and in the UV region, losses are small compared to silver and gold. Therefore, in aluminum, this SPP dispersion branch can be observed.

Since the SPP dispersion curve 6 in Fig. 4b is near the second band gap, the wavelength of the SPP should be determined by the period of the surface structure. Indeed, this wavelength is related to the propagation constant $k_x \approx 2\pi / a$ as

$$\lambda_{SPP} = \frac{2\pi}{\mathrm{Re}\,k_x} \approx a. \qquad (4)$$

The smallness of the SPP wavelength implies the subwavelength confinement of the SPP. Indeed, numerical calculations presented in Fig. 5 show that the field intensity is mainly confined near the surface. Since the frequency of the SPP is equal to the light frequency in a vacuum, it is clear that the field of the SPP is confined on a subwavelength scale.

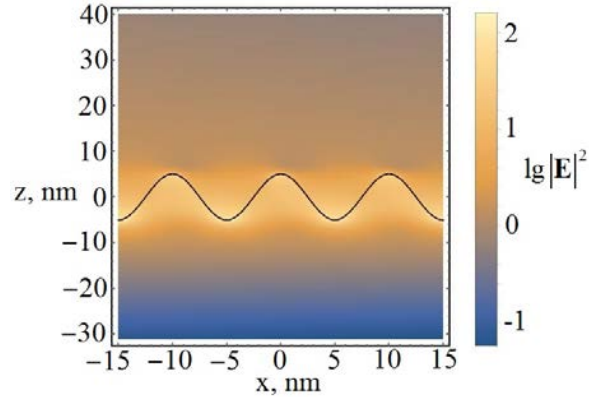

Fig. 5. The distribution of the electric field intensity $|\mathbf{E}|^2$ near the interface. The values of $\omega$ and $k_x$ used in numerical calculations are marked by the black point in Fig. 4b.

The strongly confined SPP propagates over relatively long distances. Its propagation length $l_{pr}$ can be determined by the equation

$$\frac{l_{pr}}{\lambda_{SPP}} = \frac{\mathrm{Re}\,k_x}{4\pi\,\mathrm{Im}\,k_x}, \tag{5}$$

where $l_{pr}$ is defined as the distance over which the field intensity decreases by the factor of $e$. On the sine surface with the amplitude of 5 nm, the maximum value of the SPP propagation length is about 17 SPP wavelengths. By optimizing the surface structure, one can increase this length significantly. By using the Nelder–Mead method for optimization [40], we find that the optimal surface profile is given parametrically by the equations:

$$\begin{aligned}
z(v) &= h \cdot 2^{-0.838(v/\gamma)^4}\left[1+0.075\left(2^{-14.65(v-a/3)^2/\gamma^2}+2^{-14.65(v+a/3)^2/\gamma^2}\right)-0.52\cdot 2^{-0.003(v/\gamma)^4}(v/\gamma)^2\right], \\
x(v) &= v\Big[0.996+0.024\left(e^{-10.17(v-0.13a)^2/\gamma^2}+e^{-10.17(v+0.13a)^2/\gamma^2}\right) \\
&\quad -0.16\left(e^{-10.17(v-0.26a)^2/\gamma^2}+e^{-10.17(v+0.26a)^2/\gamma^2}\right) \\
&\quad +0.16\left(e^{-10.17(v-0.35a)^2/\gamma^2}+e^{-10.17(v+0.35a)^2/\gamma^2}\right) \\
&\quad -0.29(v/\gamma)^2\left(e^{-4(v-0.083a)^2/\gamma^2}+e^{-4(v+0.083a)^2/\gamma^2}\right)\Big],
\end{aligned} \tag{6}$$

where the height is $h=18.5$ nm, the period is $a=10$ nm, $\gamma=2.5$ nm, and $v$ is the parameter varying in the range from $-a/2$ to $a/2$. The profile of the surface is shown in Fig. 6. With the surface profile given by Eq. (6), the SPP propagation length is about $53\lambda_{SPP}$ or 628 nm. Its frequency dependence is shown in Fig. 7.

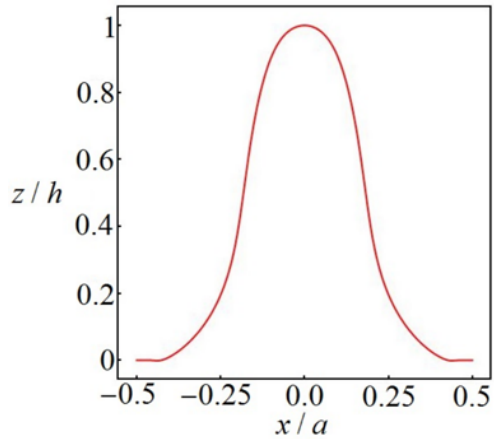

Fig. 6. The optimal surface structure defined by Eq. (6).

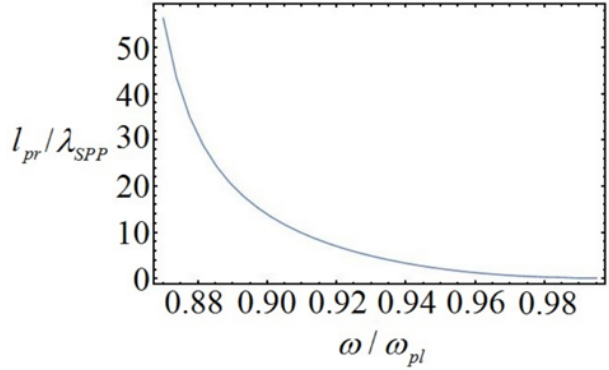

Fig. 7. The dependence of the SPP propagation length on its frequency on the optimized nanostructured interface.

## 4. MECHANISM FOR THE FORMATION OF THE STRONGLY CONFINED SPP

To explain the formation of the strongly confined SPP, let us consider the chain of spikes with the profile:

$$s(x) = \begin{cases} h\cos(2\pi x/b), & |x| < b/2, \\ -h, & b/2 < |x| < a/2, \end{cases} \quad (7)$$

where $2h$ is the height of a spike, $b/2$ is its width, and $a$ is the distance between centers of neighboring spikes. Electric charge accumulates on inhomogeneities with a small radius of curvature increasing the electric field strength near the spike [5]. Spikes, therefore, have high polarizabilities. In addition, the high polarizability of a spike can arise due to localized plasmon resonances [41-43]. If spikes have the same shape and are located close to each other, then a propagating mode arises [16].

Assuming that the chain of spikes is equivalent to the chain of two-dimensional dipoles with the polarizability $\alpha(\omega)$, one can obtain the dispersion law for the propagating mode. For the sake of specificity, we assume that a dipole moment of a spike is directed along the SPP propagation direction, $x$. The dipole moment of the $i$-th spike is determined by the equation:

$$d_i = \alpha(\omega) \sum_{n \neq i} E_{x,n}(x_i), \quad (8)$$

where $E_{x,n}(x_i)$ is the electric field of the $n$-th dipole acting on the $i$-th dipole. It can be determined by using the equation:

$$E_{x,n}(x) = d_n \left( \frac{d^2}{dx^2} + k_0^2 \right) G(x, x_n), \quad (9)$$

where $G(x, x_i) = i\pi H_0^{(1)}(k|x - x_i|)$ is the Green function of the two-dimensional Helmholtz equation. Since the chain is periodic, the dipole moments of spikes are related via the Bloch theorem:

$$d_n = d_i e^{ik_x(n-i)a}. \quad (10)$$

Excluding $d_i$ from Eqs. (7)-(9) we obtain the dispersion law:

$$1 = 2i\pi\alpha(\omega) \sum_{n=1}^{\infty} \left( \frac{d^2}{dx^2} + k_0^2 \right) H_0^{(1)}(x) \bigg|_{x=nk_0 a} \cos(nk_x a). \tag{11}$$

Let us assume that the polarization of the surface structure is defined by Eq. (7) $\alpha(\omega)$ has the Lorentz shape:

$$\alpha(\omega) \sim \alpha_m(\omega) \frac{bh}{\varepsilon_{\max} - \varepsilon_m(\omega)}, \tag{12}$$

where $\alpha_m(\omega) = (\varepsilon_m(\omega) - 1)(4\pi)^{-1}$ is the polarizability of a metal and $\varepsilon_{\max}$ is the permittivity for which the polarizability reaches its maximum value. The dispersion curves calculated by using the coordinate transformation [35,36] and with the help of Eqs. (7) and (9) are shown in Fig. 8. For calculations, we assumed that $a = b = 10$ nm, $h = 5$ nm, and $\varepsilon_{\max} = -0.5\varepsilon_d$. Both curves are qualitatively the same: they have the same slopes and are approximately in the same frequency region. This shows that strongly confined SPP modes arise due to the high polarizability of surface inhomogeneities.

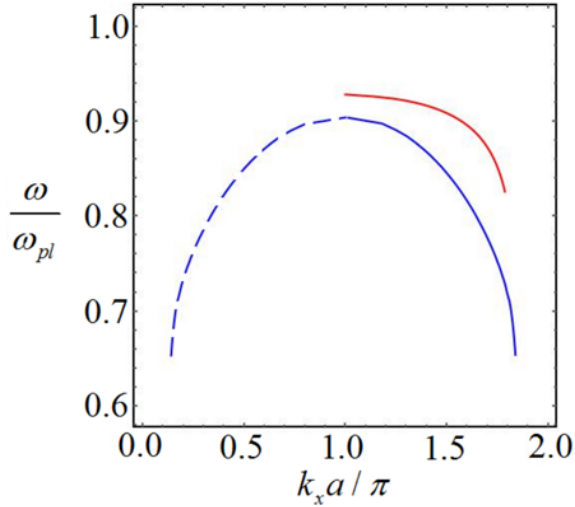

Fig. 8. The dispersion curves calculated by using the coordinate transformation [35,36] (the red line) and with the help of Eqs. (11) and (12) (the blue line). The latter curve is extended to the second Brillouin zone.

## 5. CONCLUSION

In this paper, we have shown that highly confined SPPs traveling dozens of wavelengths may exist in the far-ultraviolet region ($\lambda \sim 80$ nm). We demonstrate that SPP modes, which are strongly confined in the direction perpendicular to the direction of propagation, could arise on a periodically nanostructured interface between a metal and dielectric in the frequency range, in which $\operatorname{Re}\varepsilon_m(\omega) > -\varepsilon_d$. In aluminum, this inequality is fulfilled in the ultraviolet part of the spectrum. For these frequencies, there are no SPPs travelling along *flat* surfaces. The SPP can propagate on distance of several dozen SPP wavelengths. On the nanostructured interface considered, the modes arise due to tunneling between neighboring inhomogeneities similar to a chain of plasmonic particles. As a result, the wavenumber of the SPP is determined by the period of the nanostructure, which may be 10-20 nm. A high value of the SPP wavenumber results in strong confinement, which is crucial for SPP sensing and enhancement of nonlinear effects. Strong field localization also makes possible further miniaturization of a variety of "plasmonic optical" devices.

## ACKNOWLEDGEMENTS

A.A.L acknowledges the support of the National Science Foundation under Grants No. DMR-1312707.